\documentclass[aps,showpacs]{elsarticle}

\usepackage{amsmath,amssymb,graphicx,float,color}

\usepackage{amsmath}
\usepackage{xspace}
\usepackage{appendix}
\usepackage{float}
\usepackage{algorithm2e}
\usepackage{url}

\newcommand{\xx}{\tilde{x}}

\graphicspath{{figures/}}

\begin{document}

\begin{frontmatter}

\title{Strategic allocation of flight plans:
 an evolutionary point of view}



\author[label1]{G. Gurtner}
\author[label2]{F. Lillo}

\address[label1]{Department of Planning and Transport, University of Westminster,
35 Marylebone Road, London NW1 5LS, UK}
\address[label2]{Scuola Normale Superiore
Piazza dei Cavalieri 7,
56126 Pisa, Italy}

\begin{abstract}
We consider the simplified model of strategic allocation of trajectories in the airspace presented in \cite{jstat}. Two types of companies, characterized by different cost functions, compete for allocation of trajectories in the airspace. We study how the equilibrium state of the model depends on the traffic demand and number of airports.
We show that in a mixed population environment the equilibrium solution is not the optimal at the global level, but rather than it tends to have a larger fraction of companies who prefer to delay the departure time rather taking a longer routes. Finally we study the evolutionary dynamics investigating the fluctuations of airline types around the equilibrium and the speed of convergence toward it in finite populations. We find that the equilibrium point is shifted by the presence of noise and is reached more slowly.
\end{abstract}


\end{frontmatter}

%\linenumbers

% ===================================================================== %
\section{Introduction}
\label{section:introduction}
% ===================================================================== %

Transportation systems have a crucial importance for countries because of their social and economical impacts. The air transportation in particular is closely linked to the economical development of the areas in which it unfolds. This is why it is very important for policy makers to ensure a smooth development, even -- and especially -- in areas where the traffic increase forecasts are the highest. Indeed, the air traffic system will get closer and closer to its actual capacity, especially in Europe and in the US where the traffic could increase by 50\% in the next 20 years \cite{challenges}. As a consequence, it is important for the air traffic management world 1) to forecast the consequences on the current infrastructures and procedures and 2) to find the appropriate solutions to cope with the increase. For this reason, large investment programs like SESAR in Europe and SingleSky in the USA have been launched. 

Apart from airport capacity, one of the important bottlenecks for the increasing traffic flow will be the sectors, where the controller needs to actively separate flights in order to avoid conflicts. However, solving conflicts in areas of high traffic complexity is a demanding task, already nowadays. With the increase in traffic, the cognitive capacities of air traffic controllers will likely reach their limits and drastically increase the number of conflicts or cap the capacities of the sectors. As a consequence, navigating through the European sky will become more and more difficult in the future and will require more careful planning capacities for the network manager and for the airlines. In other words the airspace is becoming a scarce resource, especially in congested situations, like, for example, during major shutdowns of large areas (extreme weather, strikes, volcano eruptions, etc).

It is thus expected that the airlines will compete fiercely for two of the most important resources: time and space. More specifically, it is foreseen that the allocation of slots at the airports will change and include market-driven elements like bids. On the other end, the airspace will be more densely populated and the airlines will also have to compete for it. From the point of view of the  transportation companies, this increases the effort required to find better route allocation strategies, whose success depends, among other things, on the strategies adopted by the other users.

Therefore in this paper we consider the allocation of the flight plans on the airspace from the point of view of the dynamics on a complex network. This point of view is fruitfully used in different fields, like dynamics of epidemiology, information propagation on the Internet, percolation, opinion spreading, systemic risk, etc \cite{boccaletti, caldarelli}. Recently, an increasing attention is being devoted to the network description of transport systems \cite{helbing}, in particular the air transport system  \cite{li,guimera,colizza,lacasa,zanin,cardillo,plos,Sun}; for a recent review see \cite{zaninlillo}. The present study follows this stream of literature, which allows to use powerful tool to extract the main characteristic of a system, regardless on the specific details of the real system.

More specifically, we present here a simplified model based on the agent-based paradigm, particularly well suited to the problem \cite{Chakraborti}. The model describes the strategic allocation of flight plans on an idealized airspace, considered as a network of interconnected sectors. The sectors are capacity-constrained, which means that the companies might not get their optimal solutions regarding the flight plans. Thus, they will fall back to suboptimal solutions, for which they will develop different strategies. By using two different strategies for companies, we show how different factors explain the satisfaction  of different types of company. Some of these factors (the network topology and the departing times) can be regulated externally by the policy maker, while others (the mixing composition) depend on the airline population and market forces. We then study the evolutionary dynamics of the populations by considering a ``reproduction'' rate based on their past satisfaction, which acts as a fitness function. We show that this dynamics exhibits an equilibrium point which is distinct from the optimal point for the system. Moreover, the fluctuations around the equilibrium and the convergence time may hinder the convergence in practice. 

The paper is organized as follows. In section \ref{section:model}, we present briefly the model. In section \ref{section:one-route}, we present the conclusions that we drew using an earlier version of the model with only one route (two airports) and no dynamics. In section \ref{section:static}, we present some new results on the static equilibrium of the model, concerning  (i) the behaviors of the airlines in different situations (ii) the effect of the infrastructure, i.e. the number of airports and in section \ref{section:evolution} we investigate the population dynamics in an evolutionary environment. Finally, we draw some conclusions in section \ref{section:conclusions}.

% ===================================================================== %
\section{Presentation of the Model}
\label{section:model}
% ===================================================================== %

In this section we present the agent based model. We will give a brief description, whereas more details can be found in \cite{jstat} where the model has been introduced. The implementation of the model is open and can be freely downloaded for any non-commercial purpose \cite{code2}. A more detailed version of the model with a tactical part is also available \cite{tactical, code}.

The model describes the strategic allocation of trajectories in the airspace. Mimicking what is done in the European airspace, the model considers airlines who submit their flight plans to the network manager (NM). The NM checks whether accepting the flight plan(s) would lead to a sector capacity violation. If this is not the case the flight plan is accepted, otherwise it is rejected and the airline submits the second best flight plan (according to its utility function). The process goes on until a flight plan is accepted or a maximal number of rejected flight plans is reached and the flight is canceled. 
The NM keeps track of the allocated flights and checks violations of newly submitted flight plans responding in a determined way to the requests, without making counter-propositions.   

{\bf Airspace.} The airspace is modeled as a network of sectors. Topological properties of the real networks of sectors have been investigated in \cite{plos}. Each sector has a capacity, here fixed to 5 for all sectors. Some of these sectors contain airports (see below for more details on our choices) and the geometry of a flight plan is a path connecting two airports. A flight plan specifies also the departing time (see next point).  

{\bf Airlines.} The main agents of the model are the Airline Operators (AOs) who try to obtain the best trajectories for their flights and the Network Manager who accepts or rejects the flight plans. 
In the simplified version we assume that the quality of a trajectory depends on its length (the shorter, the better) and the discrepancy between the desired and actual departing/arrival time (the smaller, the better). Companies might be different depending on the relative weight of two components in their cost or utility function. Companies caring more of length are called of type ``S'' (for shifting) companies, since when their flight plan is rejected by the NM, they prefer to delay the flight but keeping a short trip length. Companies caring more of departure punctuality are called of type ``R'' (for rerouting), since when the flight plan is rejected they prefer to depart on time even if they need to use a longer route to destination. Note that in the following the AOs have only one flight. Hence in our model the optimization takes place after the previous, bigger strategic allocation of flights where AOs decides or not to operate the route, with which aircraft, etc. For this reason, all the optimization here are independent from each other for each flight, except through the capacity constraints on the network.

More quantitatively, for each flight an AO chooses a departing and arrival airport, a desired departing time, $t_0$, and selects a number $N_{fp}$ of flight plans. The $k-$th flight plan, $k=1,...,N_{fp}$, is the pair $(t_0^k, {\bf p}^k)$, where $t_0^k$ is the desired time of departure and ${\bf p}^k$ is an ordered set containing the sequence of sectors in the flight plan. 
The flight plans are selected by an AO according to its cost function. In our model it has the form 
\begin{equation}
c(t_0^k,{\bf{p}}^k) = \alpha {\cal L}({\bf{p}}^k) + \beta (t_0^k-t_0),
\end{equation}
where ${\cal L}( {\bf{p}}^k)$ is the length of the path on the network (i.e. the sum of the lengths of the edges followed by the flight). We also assume that flights are only shifted ahead in time ($t_0^k \ge t_0$) by an integer multiple of a parameter $\tau$ which is taken here as 20 minutes (all duration in this article are in minutes unless specified otherwise). The parameters $\alpha$ and $\beta$ define the main characteristics of the company. Given the discussion above, companies R have $\beta/\alpha\gg 1$, while companies S have  $\beta/\alpha\ll 1$.

{\bf Departing waves.} As it was shown in \cite{jstat}, an important determinant of the allocations is the desired departing time $t_0$ chosen by the AO. 
%Mimicking reality, we assume that within a day there are 
We assume that departing times are drawn from a distribution inside the day characterized by a certain number of peaks or waves. 

\begin{figure}[htbp]
\begin{center}
\includegraphics[width=0.48\textwidth]{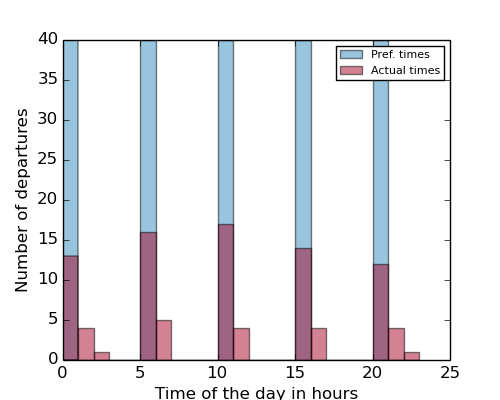}
\caption{Example of pattern of departure times (departing waves) with $N_p=5$ peaks ($\Delta t=4$) and $f_S=0.4$. The blue bars are the desired departing times while the red bars are the actual departing times after the allocation by the NM has been done.}
\label{fig:waves}
\end{center}
\end{figure}

We define first $T_d$ as the length of the ``day'' (in minutes), i.e. the time window of departure for all flights. In this time window, we define $N_p$ peaks of 60 minutes, by setting a time $\Delta t$ between the end of the peak and the beginning of the next one (thus, $N_p = T_d/(\Delta t + 60)$). The parameter $T_d$ is fixed to 24 hours in the following. Figure \ref{fig:waves} shows an example of departing waves.
%Then, we define either a total number of flights and divide them equally between peaks (within which they depart at random), or we set a given number of flights per peak.
Then we define a total number of flights $N_f$ and divide them equally between peaks. In the following, we also use the corresponding time density $d = N_f/24$, i.e. the average number of flights per hour.

{\bf Dynamics.} Given a mixed population of AOs of different types, at each time step, an AO is selected randomly\footnote{The random order of arrival of bids is chosen to guarantee that neither type of company has an advantage because it arrives first to the network manager.}. The AO chooses the departing and arrival airports and the desired departing time $t_0$ for its flight, drawing it from the departing time distribution. It then computes the $N_{fp}$ best flight plans for the flight according to its cost function and submits them, one by one in increasing order of cost, to the NM. The NM accepts the first flight plan which does not cross overloaded sectors, i.e. at already maximal capacity. If none of the  $N_{fp}$ flight plans is accepted, the flight is rejected and the satisfaction of the AO for this flight is set to 0.

{\bf Metrics.} The metric measuring the satisfaction (or fitness) of a company about a given flight $f$ is
\begin{equation}
{\cal S}_f=c_f^{best}/c_f^{accepted},
\end{equation}
where $c_f^{best}$ is the cost of the optimal flight plan for the flight $f$ according to the AO cost function (i.e. the first flight plan to be submitted for the flight), and $c_f^{accepted}$ is the cost of the flight plan eventually accepted for this flight. If no flight plan has been accepted, we set ${\cal S}_f$ to 0. Note that ${\cal S}_f$ is always between 0 and 1. The value 1 is obtained when the best flight plan is accepted.

Since the AOs have only one flight here, the satisfaction of a flight is also the satisfaction of its company. When several companies have the same type (same ratio $\beta/\alpha$), we make use of the average satisfactions across them. For instance, $\mathcal{S}^S$ and $\mathcal{S}^R$ are respectively the average satisfactions of companies S with $\beta/\alpha\ll 1$ and R with $\beta/\alpha\gg 1$. Finally, we use also the average satisfaction across all flights as a measure of the global satisfaction of the system:
%We then consider the average satisfaction of a company as the sample average of the satisfaction across all the flights of the company.
%When more than one AO is present in the system, the definition of the satisfaction for the whole system is not unique and more complex.
\begin{equation}\label{eq:GS}
{\cal S}^{TOT} = f_S \times {\cal S}^{S} + f_R \times {\cal S}^{R},
\end{equation}
where $f_i$ and ${\cal S}^{i}$ are the fraction of flights and the average satisfaction of company $i$, respectively, and $f_S+f_R=1$.

The simulations we describe hereafter are obtained for  type of airspace different from the one used in Ref.\cite{jstat}. In order have a more controlled environment, we generate a triangular network with 50 nodes. Each node of the network represents the center of an hexagonal sector. Sectors are linked to each of their neighbors. In order to avoid paths having exactly the same duration, which could lead to ties in the optimization, we sample the crossing times between sectors from a log-normal distribution so as to have a 20 minutes average and a very small variance (inferior to $10^{-4}$ minutes). 

The number of airports available to the air companies can be chosen before the simulation. Unless specified otherwise, we fix the number of airports to 5. In the following, we present results in which we drew 10 times the airports randomly, then ran 100 independent simulations on each of these realizations.

The previous setup is very stylized but allows to catch the main features of the model. In \ref{annex:rob} we present some  robustness checks of the model by considering two more realistic setups. In the first one we consider a scale free network of airports, i.e. not all the airports are equivalent in terms of number of flights/destinations, but hubs and spokes are present. In the second we use real ECAC data to construct the network of sectors, the sector of capacities, the origin/destination frequencies, and the wave structure. We find that the results are indeed similar to the ones presented thereafter for the stylized model, where a much more controlled setting is used.

% ===================================================================== %
\section{Competition over a single route}
\label{section:one-route}
% ===================================================================== %

Ref. \cite{jstat} considered the model in a static setting with only two airports. Static means that no evolutionary dynamics has been considered. Ref. \cite{jstat} showed that with a single type of company, there is a (congestion) transition, much like the congestion observed in other transport systems, e.g. car traffic, when the number of flights becomes too large. 
When two extreme types of companies (R and S) are competing for the airspace, ref. \cite{jstat} showed that there exists a unique fraction of mixing corresponding to a stable Nash equilibrium. The strategies are interacting positively, leading to an absolute maximum in satisfaction for the overall system at a mixing fraction different from 0 and 1.
Finally, by performing extensive simulations, ref. \cite{jstat} showed that overlaps between different possible paths connecting two airports is an important determinant of the possibility to gain advantage from a rerouting, impacting directly the satisfaction of companies R and indirectly the one of companies S.

Compared with \cite{jstat}, the main innovations presented here are:
\begin{itemize}
\item We consider a more realistic setting with multiple airports and we study the relation between number of flights and number of airports such that the satisfaction of airlines is preserved.
\item We study the specific mechanism linking the structure of the paths on the network, the overlap, to the increase in traffic on this network.
\item We consider an evolutionary setting where the capability of a type of company of continuing its business depends on the past satisfaction. 
\end{itemize}

% ===================================================================== %
\section{Static equilibrium}
\label{section:static}
% ===================================================================== %

% --------------------------------------------------------------------- %
%\subsection{Static Equilibrium}
%\label{subsection:static}
% --------------------------------------------------------------------- %

We first compare the general output of the simulations to the case where there is only one connection and two airports. For this, we study directly the case where we have two different types of company, S and R, which are competing on the airspace. Note that each company has only one flight, whose pair origin/destination is drawn randomly from the available ones.

\subsection{Departing patterns and mixed populations}
\label{subsubsection:traffic}

Our first aim is to understand how the satisfaction of each type of company depends on its environment. For this, we fix the number of flights (to $N_f = 24 \times d = 480$ here) and we change the proportion $f_S$ of companies of type S present in the airspace. This last parameter will be called mixing parameter in the following. We also change the structure of the wave pattern, by changing the parameter $\Delta t$, defined previously.

\begin{figure}[htbp]
\begin{center}
\includegraphics[width=0.48\textwidth]{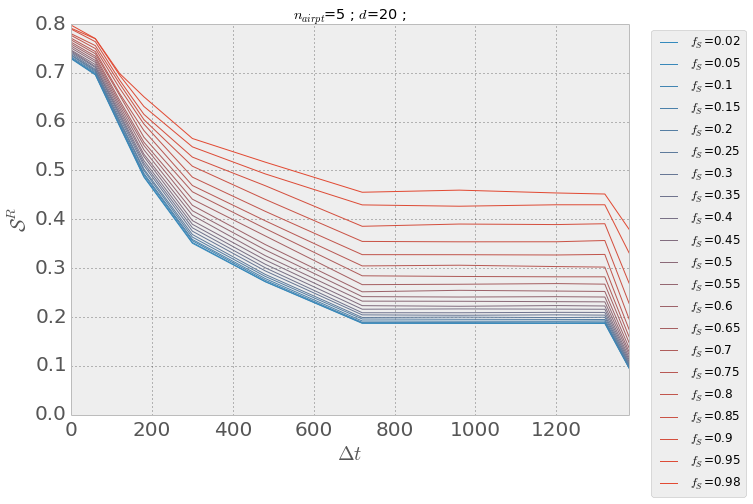}
\includegraphics[width=0.48\textwidth]{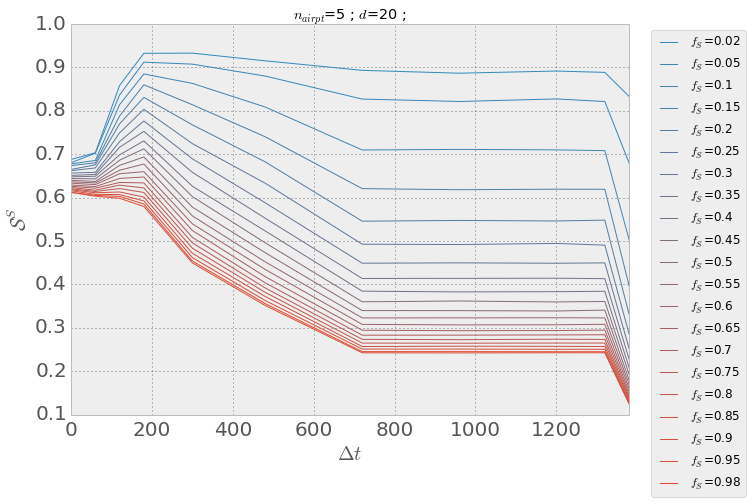}
\caption{Satisfaction of companies as a function of the time $\Delta t$ between waves. Different lines refer to different mixing parameters. The number of airports is 5 and the number of flights is $20 \times 24$. Left: satisfaction of companies R. Right: satisfaction of companies S.}
\label{fig:sats_vs_dt}
\end{center}
\end{figure}

\begin{figure}[htbp]
\begin{center}
\includegraphics[width=0.48\textwidth]{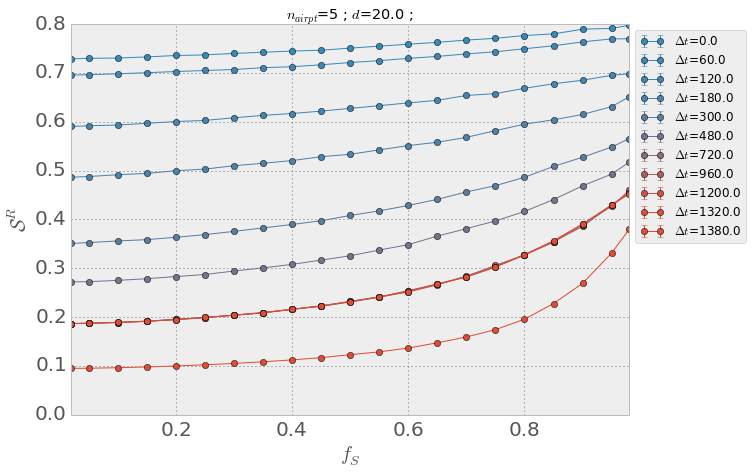}
\includegraphics[width=0.48\textwidth]{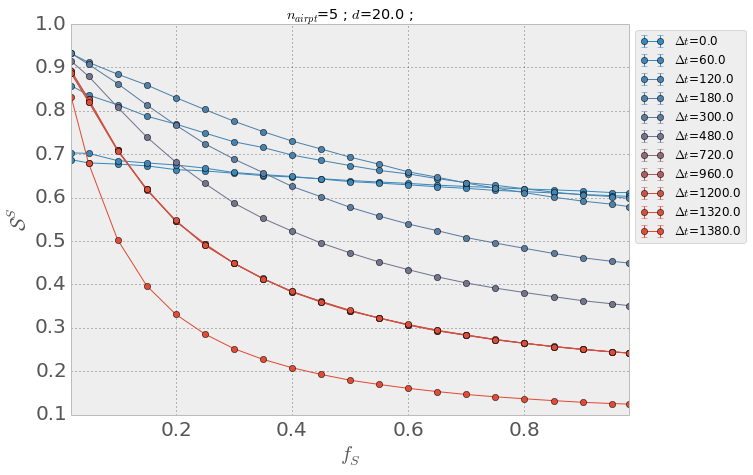}
\caption{Effect of the mixing parameter on the satisfaction for different wave patterns. The number of airports is 5 and the number of flight is $20 \times 24$. Left: satisfaction of companies R. Right: satisfaction of companies S.}
\label{fig:sats_vs_fs}
\end{center}
\end{figure}

Figures \ref{fig:sats_vs_dt} and \ref{fig:sats_vs_fs} show the satisfaction of the two types of company as a function of $\Delta t$ for different values of the mixing parameter.  The results for companies R are quite intuitive. These companies are better off when they are competing with a large fraction of companies S (figure \ref{fig:sats_vs_fs} left) and when there are more waves, i.e. when $\Delta t$ is small (figure \ref{fig:sats_vs_dt} left). This is expected, since more waves means more ``space'' for companies when the number of flights is fixed. Moreover, companies R have a stronger dependency on the mixing parameter when the number of waves is small, i.e. $\Delta t$ is big. In other word, they compete more strongly with each other in this case. Note that the plateau present for $\Delta t \in [700, 1300]$ is due to the fact that for this range of parameter, the number of waves is constant and they are far from each other. Indeed, for this range, there are only two peaks, which come slowly apart as $\Delta t$ increases. Since they are sufficiently apart, the flights from the previous wave do not interact with the flights in the next one, and the satisfaction do not change with $\Delta t$. In other words, the first flight from the second wave departs after the last flight from the first wave arrives.

The effect of the environment is more complex for companies S. Indeed, for some values of the mixing parameter $f_S$, their satisfaction is not monotonous with $\Delta t$. This behaviour is explained by the following trade-off. On one hand, the bigger $\Delta t$ is, the less waves there are. Hence companies are competing effectively with a higher number of other companies, which is the reason behind the decreasing curve of companies R in the left panel of figure \ref{fig:sats_vs_dt}. On the other hand, companies S have a strategy where they try to delay their flight if the first flight plan is rejected. This means that if the waves are too close to each other, the delayed flight plans will likely conflict with the flights in the next wave. For this reason, their satisfaction increases at the beginning when the waves comes apart and then decreases when the waves are further apart and in smaller number. 

Note that first effect -- the decrease of satisfaction due to concentration within waves -- is less significant when companies S compete with many R companies. Indeed, in this case, their increasing concentration within a wave is of little importance for them, because they can always shift their flight plan two or three times to get out of the wave and not conflict anymore with companies of type R. For this reason, their curve is monotonous with $\Delta t$ for high values of the mixing parameter. More strikingly, their satisfaction is higher for very high $\Delta t$ than for very small ones if $f_S << 1$.

Hence, the companies are reacting differently to different environment (waves) because they are sensitive to different mechanisms. The interplay of the mechanisms lead to interesting patterns that translate in interesting behaviours when framing the model in an evolutionary environment. But before coming to this, we inspect in the following section the effect of the density of airports on the companies.

\subsection{Effect of the density of airports}
\label{subsubsection:infrastucture}

The increase of the number of airports in real airspaces has some obvious impact. In our model, for a fixed number of flights, increasing the number of airports leads to more potential routes and thus less interactions between flights. However, it is not clear whether this effect is similar to a decrease of the number of flights with a fixed number of airports. In order to investigate this problem, we repeat some simulations with constant parameters $\Delta t = 60 \times 5$ and $f_S = 0.5$ but with different number of flights and different number of airports. 

\begin{figure}[htbp]
\begin{center}
\includegraphics[width=0.48\textwidth]{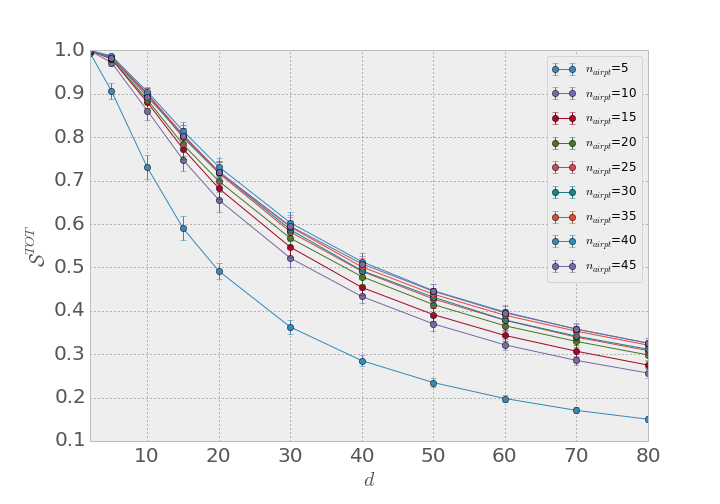}
\includegraphics[width=0.48\textwidth]{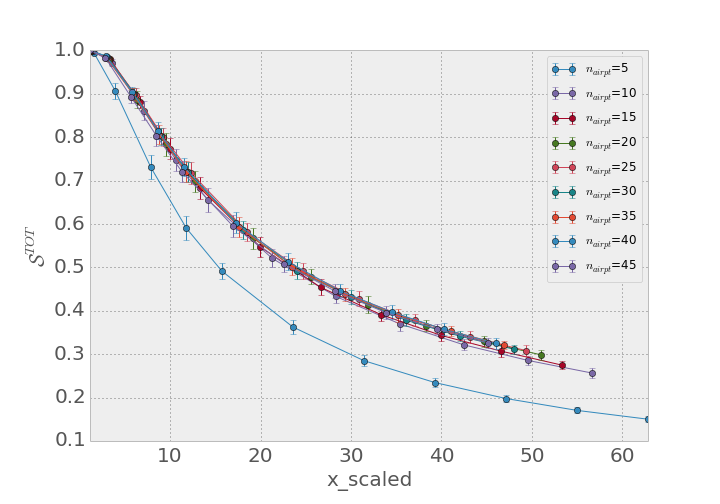}
\caption{Satisfaction against density for different number of airports. Bother parameters seem to have opposed effects. Left: non-rescaled plot. Right: the abscissa is rescaled by $d/n_{airpt}^{0.15}$.}
\label{fig:airport_sweep}
\end{center}
\end{figure}

The results are presented in figure \ref{fig:airport_sweep}. As one can see on the left panel, the average satisfaction decreases with the number of flights, and increases with the number of airports. In order to find a relationship between both parameters, we rescaled the abscissa by $d/n_{airpt}^{\alpha}$, trying to find the value of $\alpha$ where the curves would collapse the best. Purely empirically, we found that $\alpha \simeq 0.15$ is the best match that we could get, except for very low numbers of airports, for which the curve does not collapse well with the others (see right of figure \ref{fig:airport_sweep}). We do not have an analytical argument to ground this scaling, but we suspect that it is linked to the degree of the network, since $\alpha = 0.15 \simeq 1/6$, and 6 is the degree of the triangular lattice on which the airspace is embedded.

Whatever the reason is behind the exact scaling, it is thus obvious that both parameters play some inverted roles. Roughly speaking, more airports give more choices to companies, and more flights ``fill'' these choices. In order to captures this point, we computed a metric $Q$ that we call ``overlap'' and which represents how much the paths open to companies are similar to each other. More specifically, if $p_1$ and $p_2$ are two paths on the network, we compute first:
$$
Q_{p_1, p_2} = \frac{p_1 \cap p_2 }{p_1 \cup p_2},
$$
which is simply the number of common nodes (sectors) in $p_1$ and $p_2$ divided by the total number of unique sectors in $p_1$ and $p_2$. To compute an aggregated value, we consider all flight plans computed by the companies, i.e. including also the suboptimal ones. From them, we consider all the paths contained in the flight plans, and we compute the overlap between ALL the pairs of possible paths to obtain $Q$. Note that this metric does not consider the time at all, so it might be that two flight plans with the same path actually departs at very different times and have no chance of interacting. This could be called a ``static'' overlap, but we consider this metric because it is simple and it is very specific to the network, rather than the companies themselves. Even with this simple metrics, one can catch an interesting feature of the model.

\begin{figure}[htbp]
\begin{center}
\includegraphics[width=0.48\textwidth]{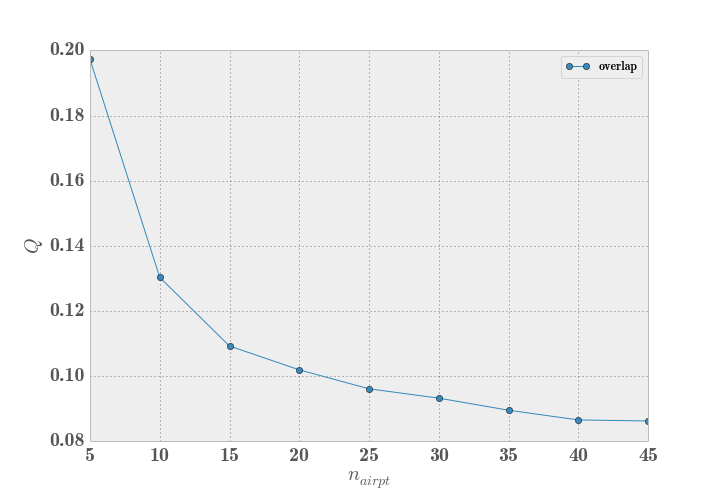}
\includegraphics[width=0.48\textwidth]{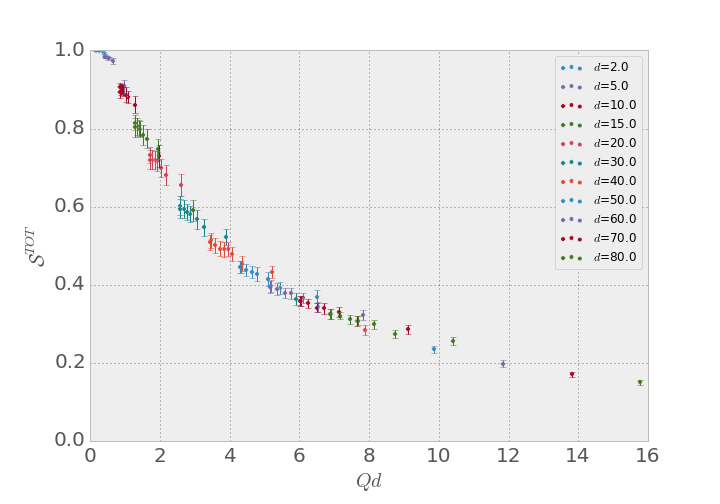}
\caption{Left: Overlap between paths as a function of the number of airports in the airspace. Right: Total satisfaction against $Q\cdot d$ for different densities and overlap. The curves collapse very well.}
\label{fig:airport_sweep_overlap}
\end{center}
\end{figure}

Left panel of figure \ref{fig:airport_sweep_overlap} shows the overlap as a function of the number of airports in the airspace. As expected, the overlap between potential paths decreases with the number of airports. The overlap is a very physical quantity, which have a tight connection with the number of flights. Indeed, when one opens a path on the network, in average to capacities of the other paths to accept flights decrease by $Q$. So the ``density'' of flights per route is effectively $Q N_f/N_{routes}$, where $N_{routes}$ is the total number of routes. Since the average satisfaction is likely to be a function of this density, all curves for different number of flights and different number of airports should scale as $Q d$. 
This is exactly the result we obtain in the right panel of figure \ref{fig:airport_sweep_overlap}, where we plot the total satisfaction against $Q d$. As expected, all the curves collapse very well.

This result shows that the effect of a change of a number of airports can be deduced from the effect of a change in traffic, or vice versa. It has a very practical impact, which is that the simulations can be run on different number of airports, or different number of flights, but not necessarily both. This is why in the following we keep the number of airports to 5, and only study the effect of variations of density. It could have also a more general impact, in the sense that if the results would hold on a more realistic airspace (which should be the case because of the general scope of the overlap metric), a policy maker could for instance try to push for the creation of new airports to counter balance increasing traffic. This, of course, suppose that the demand is constant and not too localized (i.e. an additional airport in a big city).

\subsection{Global satisfaction: equilibrium state}
\label{subsubsection:global_eq}

\begin{figure}[htbp]
\begin{center}
\includegraphics[width=0.48\textwidth]{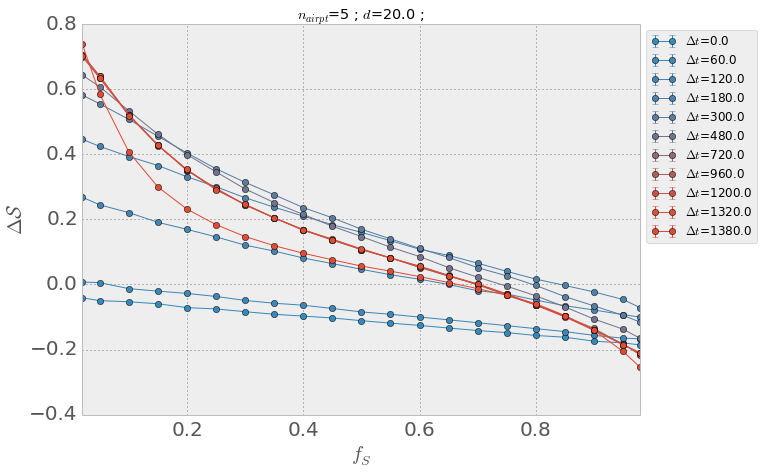}
\includegraphics[width=0.48\textwidth]{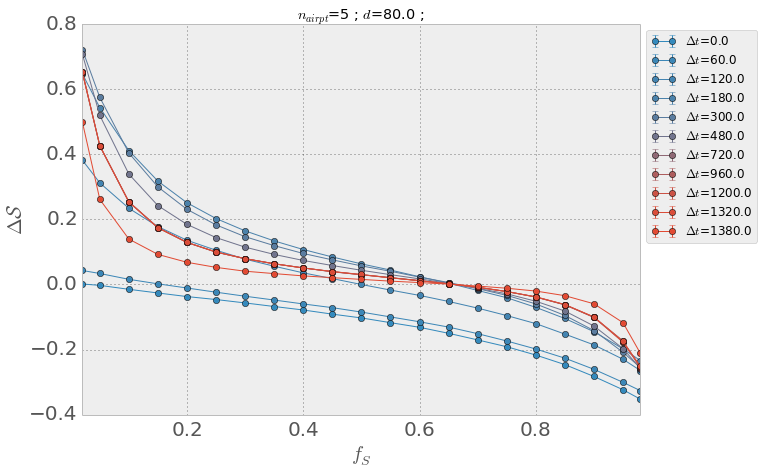}
\caption{Difference of satisfaction $\Delta S$ between companies S and companies R versus the mixing parameters for different values of $\Delta t$. Left: low density of flights, $d=20$. Right: high density of flights, $d=80$.}
\label{fig:diff_S}
\end{center}
\end{figure}

As seen in section \ref{subsubsection:traffic}, the effect of the environment properties on the satisfaction of two types of companies is different. What is not clear yet is the exact interplay of the mechanisms and the resulting difference of satisfaction of the populations. In figure \ref{fig:diff_S} we plot the difference of satisfactions $\Delta S$ between population S and population R, as a function of the mixing parameter $f_S$ and for several values of $\Delta t$. In the left panel, the density of flights is quite small ($d=20$), corresponding to the one used in figures \ref{fig:sats_vs_dt} and \ref{fig:sats_vs_fs}. On the right panel we show the result for a much higher density ($d=80$), corresponding to a congested airspace.

The difference of satisfaction $\Delta S$ between the two populations depends heavily on the two parameters. At both densities, the first values of $\Delta t$ are clearly crippling population S, since in this configuration $\Delta S$ is always negative. This is due to the fact that very frequent waves prevent companies S to delay their flight, whereas companies R can find available path by suitable rerouting. For higher values of $\Delta t$, the situation becomes more favourable to companies S, since the difference is usually positive. The details of the variations of the difference are quite complex with the two parameters, but it is clear it is always decreasing monotonically with $f_S$. The point where it crosses 0 varies with $\Delta t$ but not wildly (except for small $\Delta t$).

It is worth noting that these curves can be considered as fitness curves for two populations competing for the same resources in a given environment. If the higher fitness affects positively future reproduction rate (i.e. the possibility of continuing and expanding business), we show in section \ref{section:evolution} how to study the dynamics of the two populations in an evolutionary framework. Here we simply recall that the points where the difference of fitness curves vanish (i.e. its roots) are equilibrium points for the dynamics.  The existence of a single root (as in Fig. \ref{fig:diff_S}) shows that there is only one equilibrium point (a part the two absorbing states at $f_S=0$ and $f_S=1$). The slope of $\Delta S$ at its root measures the stability. Since the slope we observe is negative, the equilibrium is stable. In other words, when the proportion of companies S is too high, their satisfaction/fitness decreases, thus giving a lower reproduction rate for them, favoring companies R, and driving back the system towards the equilibrium.  

\begin{figure}[htbp]
\begin{center}
\includegraphics[width=0.48\textwidth]{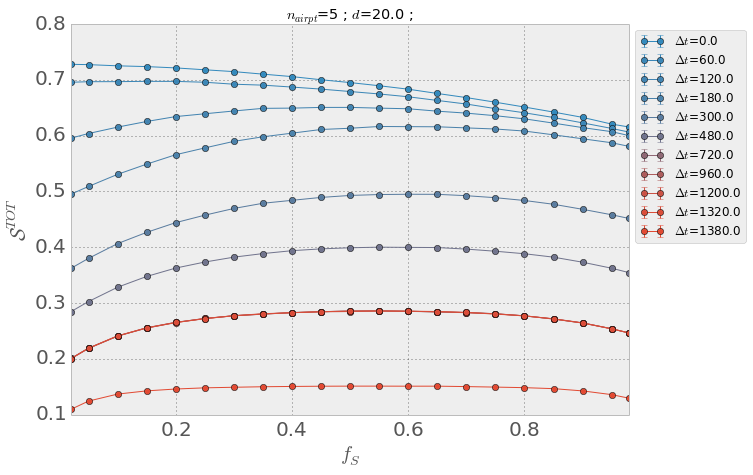}
\includegraphics[width=0.48\textwidth]{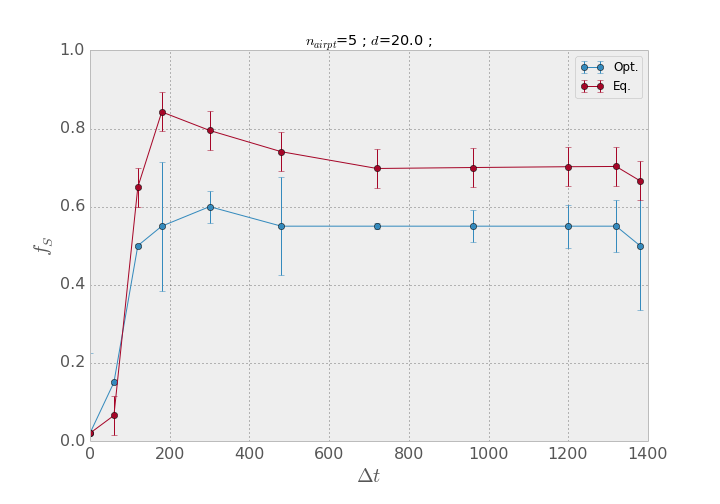}
\caption{Left: average satisfaction of all companies against the mixing parameters for different wave patterns. Right: evolution of the maximum satisfaction and the equilibrium point as a function of $\Delta t$.}
\label{fig:global_opt}
\end{center}
\end{figure}

Another important question in this kind of system is whether the equilibrium point is optimal also for the system. For this reason we compute also the global satisfaction, Eq. \ref{eq:GS}, which is the average satisfaction of all the flights. A higher global satisfaction means that globally the system is in a better shape, leading to increased profits for airlines and possibly better service for passengers. In the left panel of figure \ref{fig:global_opt} we show the value of the global satisfaction as a function of the mixing parameter $f_S$ for different values of $\Delta t$. The first conclusion is that the global satisfaction is usually better for $0<f_S<1$ than for pure populations. This should not be a surprise, because we saw that each population performs better against the other one. This is the typical case where different populations have different niches and thus their interaction is beneficial for both. The second conclusion is that for all values of $\Delta t$, there exists a unique maximum and its position varies with $\Delta t$.

On the right panel of figure \ref{fig:global_opt}, we plot both the global optimum, extracted from the left panel, and the equilibrium point, extracted from the left panel of figure \ref{fig:diff_S}. Both exhibit similar variations. For small values of $\Delta t$, the equilibrium point and the global optimum are both at $f_S\simeq 0$. When $\Delta t$ increases, companies S increase their advantage against companies R because they are not troubled by the next wave. Then both values decrease, to stabilize at value $f_S >0.5$, showing the greater advantage of companies S when the departing pattern is composed by well separated waves. 

More importantly, both curves are clearly distinct for $\Delta t \gtrsim 100$, even considering error bars. This is an important result, because it shows that the equilibrium mixing condition is not the optimal at the global level. In particular the evolution of the system toward its equilibrium mixing would tend to favour drastically population S, whereas the global optimum would be reached with a much smaller market share of companies S. This is exactly where policy makers should step in and issue policies driving the system to the optimum. 

%\newpage

% --------------------------------------------------------------------- %
\section{Evolutionary Dynamics and Equilibrium}
\label{section:evolution}
% --------------------------------------------------------------------- %

In the previous section, we interpreted the satisfaction of each company as its fitness when competing with the others for the same resources -- namely  time and space. Interpreting these fitnesses as the capability of expanding business, it is possible to develop a dynamical evolutionary model for studying the dynamics toward equilibrium and its fluctuations, as well as the role of finite size populations.

In our model each type of company has a population size at time $t+1$ which depends on its satisfaction at time $t$. In order to keep the simulations under reasonable computational time, and following what is done in evolutionary biology models \cite{nowak} we keep the total population fixed. This means that only the mixing parameter $f_S$ is changing between time $t$ and $t+1$. For the reproduction rule, we use an exponential reproduction, i.e. the rate of reproduction of a population is proportional to its fitness and its current population. Combined to the fixed population conditions, this leads to a discretized version of the so-called replicator model \cite{nowak}:

$$
f_S^{t+1} = f_S^{t} + \Delta S_t\, f_S^t\, (1-f_S^t),
$$
where $f_S^t$ is the mixing parameter at time $t$ and $\Delta S_t$ is the difference in satisfaction between companies S and R at time $t$. In the simulations, we also choose to always keep the number of companies of each kind to a minimum of 1. This ensures that the equilibria at $f_S = 0$ and $f_S = 1$ do not act as absorbing barriers (sinks). Indeed, since the populations are finite, a small non-null $f_S^t$ could lead to exactly 0 company S, which leads in turn to $f_S^{t'} = 0$ for all $t'>t$. Analogously the same happens when $f_S^t = 1$. Note that all other parameters ($\Delta t$, number of airports, airpspace structure, etc) are being kept constant throughout the reproduction process, i.e. the environment is stable.

\begin{figure}[htbp]
\begin{center}
\includegraphics[width=0.48\textwidth]{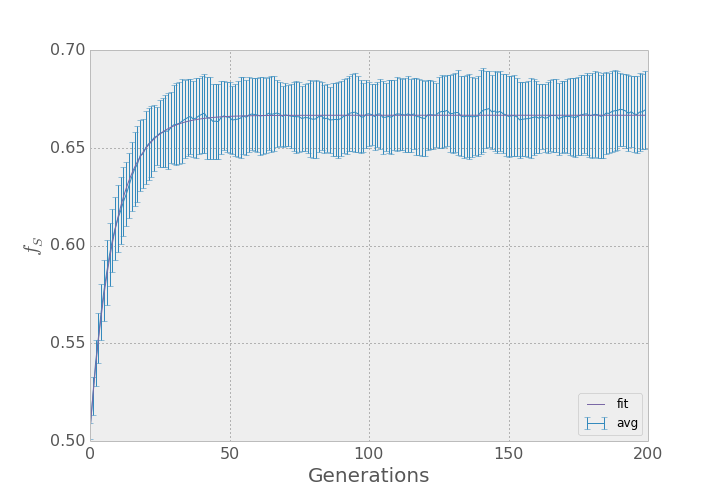}
\includegraphics[width=0.48\textwidth]{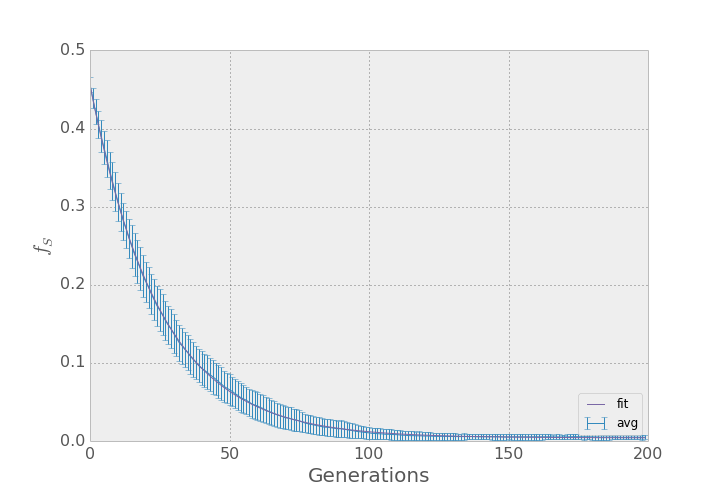}
\caption{Evolution of the mixing parameters with the generations, averaged over 100 realizations (and only one network realization). The blue lines are the averages, the violets lines are exponential fits, and the error bars are the average standard deviations. The coefficients of determination of the regressions are over 0.98. Left: $\Delta t = 23\times 60$. Right: $\Delta t = 0$. }
\label{fig:evo}
\end{center}
\end{figure}

Figure \ref{fig:evo} shows the results of the simulations for two distinct values of $\Delta t$. The plots show the evolution of the mixing parameter with time (i.e. the number of generations). The solid blue lines are averages over 100 runs and the solid violet lines represent the results of exponential fits. Both lines are well fitted ($R^2>0.98$), and the equilibrium is clearly reached in both cases. On the left, there is only one wave of departure, thus companies S have an advantage and the point of equilibrium for $f_S$ is above 0.5. On contrary, the figure on the right shows that when there are no waves ($\Delta t=0$), companies S are very disadvantaged, and the point of equilibrium is close to 0. Note that both figures are roughly consistent with figure \ref{fig:diff_S} on the left, where the difference in fitness $\Delta S$ has a root close to 0 when $\Delta t = 0$ and has a root close to 0.7 when $\Delta t = 23\times 60$. 

\begin{figure}[htbp]
\begin{center}
\includegraphics[width=0.48\textwidth]{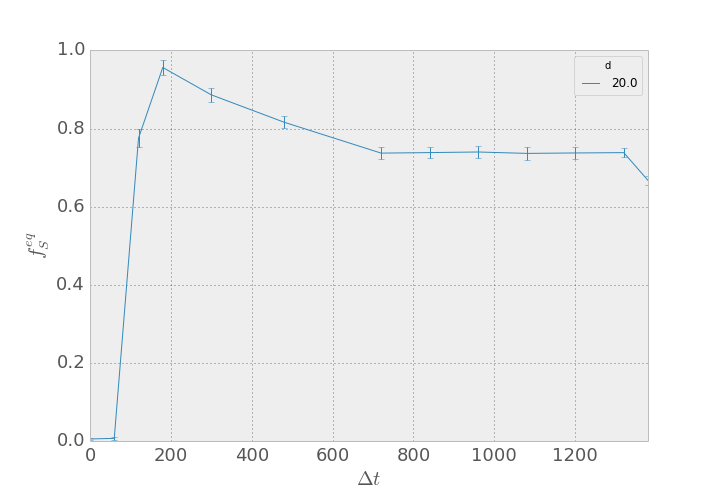}
\caption{Evolution of the equilibrium coming from the evolutionary simulations with $\Delta t$.}
\label{fig:evo_eq}
\end{center}
\end{figure}

To investigate more in detail the difference between the two results, in figure \ref{fig:evo_eq} we plot the position of the equilibrium point -- computed by averaging the last 40 generations in each run -- as a function of $\Delta t$. The plot is directly comparable to the right panel of figure \ref{fig:global_opt}, since all the other parameters are the same. The curves are roughly similar, but the one obtained with evolutionary dynamics displays larger values of $f_S$, especially around  $\Delta t = 3\times 60$, than the static one. It can be shown that the stochasticity of the fitness function can lead to such a result, see \ref{annex:sde}. This is an important results, because the noise coming from the fitness function can drive the equilibrium even further from the global optimum than in the deterministic case.

We now consider how the system converges to the equilibrium and how external parameters like $\Delta t$ influences the convergence. It is worth reminding the link between the fluctuations around the equilibrium and the shape of the fitness functions. Indeed in the continuous version of the replicator model, the stability of the equilibrium is given by the slope of the difference of fitnesses at its root \cite{nowak}. Higher absolute slopes translate into a higher stability and faster convergence to the steady state. 
%In the case where the replicator model is used in discretized times, the fluctuations around the equilibrium is governed by the same slope. 
However our system does not have a deterministic fitness function, since the satisfaction depends on the specific realization of the model. This additional noise, directly linked to the mechanisms embedded in the model, affects both the fluctuations around the equilibrium and the time of convergence in general. In \ref{annex:sde}, we briefly show analytically why this is the case. 

\begin{figure}[htbp]
\begin{center}
\includegraphics[width=0.48\textwidth]{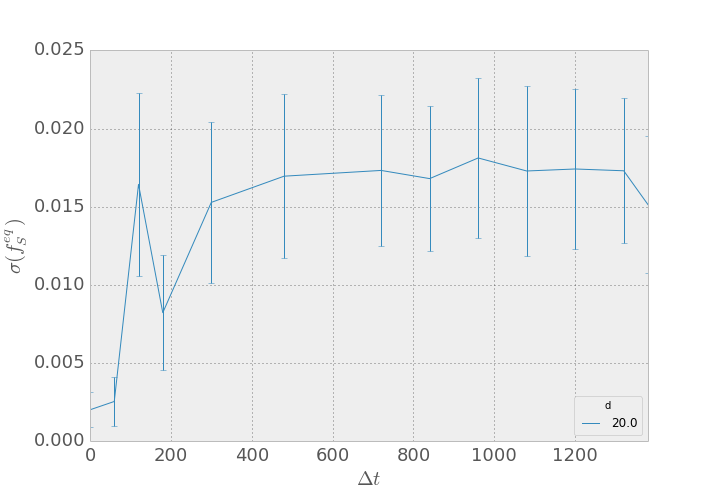}
\includegraphics[width=0.48\textwidth]{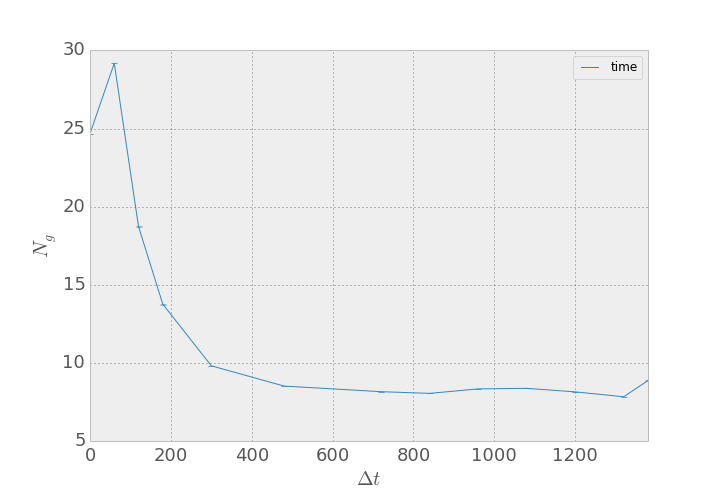}
\caption{Left: evolution of the standard deviation of the value of $f_S$ when the equilibrium is reached against $\Delta t$. Right: Typical number of generations before the equilibrium is reached (time to equilibrium) as a function of $\Delta t$.}
\label{fig:evo_std}
\end{center}
\end{figure}

The magnitude of the fluctuations around the equilibrium are depicted on the left panel of figure \ref{fig:evo_std}. There is a weak trend towards bigger fluctuations when $\Delta t$ increases but their magnitude reaches a plateau quickly. Note that the standard deviation is far from between negligible, implying that the fluctuations are typically 15\% of the value of the equilibrium point. This means that the static analysis performed in section \ref{section:static} is far from revealing all the features of the model. 

The time of convergence to equilibrium is plotted in the same figure on the right panel. This is the characteristic time obtained by fitting the $f_S^t$ with an exponential function of time. This time scale is quite high for small values of $\Delta t$ -- where the fluctuations are small -- but decreases to a small value (around 7 or 8 generations) when $\Delta t$ increases -- where the fluctuations are high. As already stated, the magnitude of the fluctuations depends on two independent mechanisms, the variance of the fitness function and its slope. In order to understand which mechanism plays a major role, we performed a regression with an ordinary least-square procedure. The dependent variable  is the inverse of the time to equilibrium, and the two explanatory variables are $\sigma_S$, the standard deviation of the fitness function around the equilibrium point, and $\gamma$, the slope of the fitness function. The results of the regression are presented in table \ref{tab:ols}.

\begin{table}
\centering
\begin{tabular}{c|cccc}
Variable 	& Weight	& Std. err.	& t-test	& Conf. Int. \\
\hline
Const		& 0.0911	& 0.023		& 0.004		& [0.038, 0.144]\\
$\sigma_S$	& -0.2588	& 0.046		& 0.001		& [-0.366, -0.152]\\
$\gamma$	& -0.1279	& 0.029		& 0.002		& [-0.196, -0.060]\\
\end{tabular}
\caption{Results of the ordinary least square regression of $1/tau$ with the estimated parameters, the standard errors, the p-values of a t-test, and the 5\% - 95\% confidence intervals. The coefficient of determination is $R^2 = 0.94$.}
\label{tab:ols}
\end{table}

\begin{figure}[htbp]
\begin{center}
\includegraphics[width=0.48\textwidth]{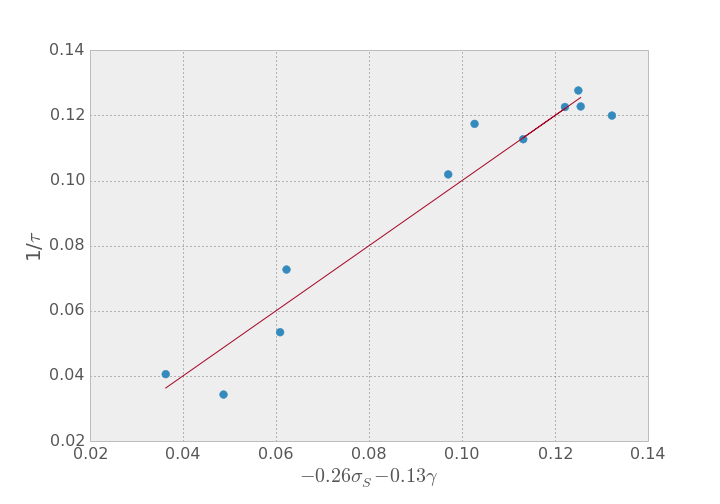}
\includegraphics[width=0.48\textwidth]{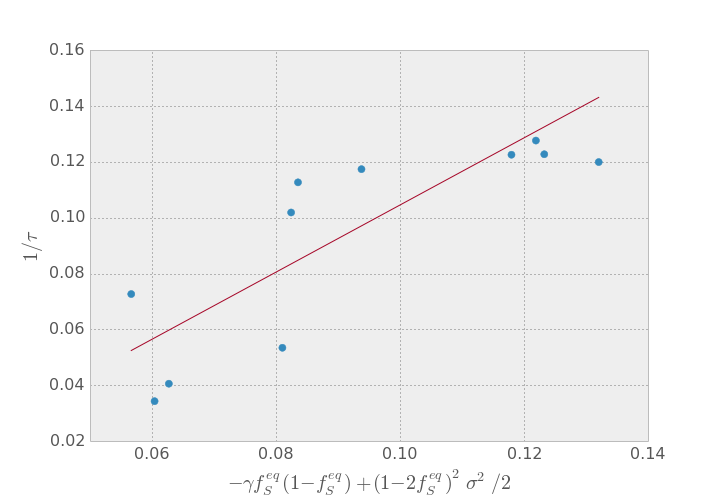}
\caption{Inverse of time to equilibrium versus a combination of the standard deviation and the slope of the fitness function. Left: the weights are the results of an ordinary least square ($R^2 = 0.94$) regression. The solid red line shows the results of the regression. Right: the weights are coming from analytical arguments. The solid red line is a linear regression ($R^2 = 0.67$).}
\label{fig:ols}
\end{center}
\end{figure}

The regression is very good, with the coefficient of determination of 0.94. Both variables impacts negatively the inverse of time to equilibrium. This is expected, since a higher variance of the fitness function should increase the time to equilibrium, as well as a higher slope (because the slope is negative). Finally, we can conclude that both mechanisms play an important role, since both coefficients are similar in magnitude. Note however that the variance of the fitness function is twice as important as the slope to determine the dynamics of the system. As a consequence, one cannot simply infer the dynamics from the static considerations made in section \ref{section:static}. Finally, we show in figure \ref{fig:ols} the graphical results of the regression, plus a plot showing the variation of $1/\tau$ against the expression found analytically (see \ref{annex:sde}). The agreement is worse than with the regression, which might be due to the fact that the analytical model is linearised around the equilibrium, whereas the system can in fact start quite far from it.

\begin{table}
\centering
\begin{tabular}{c|cccc}
Variable 	& Weight	& Std. err.	& t-test	& Conf. Int. \\
\hline
Const		& 0.0101	& 0.008		& 0.244		& [-0.008, 0.029]\\
$\sigma_S$	& -0.0338	& 0.016		& 0.071		& [-0.071, 0.004]\\
$\gamma$	& -0.0226	& 0.010		& 0.059		& [-0.046, 0.001]\\
\end{tabular}
\caption{Results of the ordinary least square regression of the fluctuations of the mixing parameter with the estimated parameters, the standard errors, the p-values of a t-test, and the 5\% - 95\% confidence intervals. The coefficient of determination is $R^2 = 0.74$.}
\label{tab:ols_std}
\end{table}

The same procedure can be used to study the fluctuations around the equilibrium. This time the variable to be explained is the standard deviation of the mixing parameter over the last 40 generations of each run (see Table \ref{tab:ols_std}).
In this case the regression is not very good, even if some variance is still explained by the variables. Strikingly, it is the signs of the coefficients which are important. For example, when the variance of the fitness function $\sigma_S^2$ is higher, the fluctuations around the equilibrium are actually smaller. Likewise, when the slope is higher (increasing towards 0), the fluctuations are smaller. This counter-intuitive results comes from the fact that the fluctuations depend also on the position of the equilibrium point, which are obviously vanishing when $f_S \rightarrow 1$ or $f_S \rightarrow 0$.
%As showed in \ref{annex:sde}, the variance of $f_S$ is bigger around $f_S = 0.5$ and vanishes for $f_S \rightarrow 0$ and $f_S \rightarrow 1$. Note that the inclusion of $f_S^{eq}$ as an explanatory variable in the regression improves slightly the results ($R^2 = 0.84$). However the relationship are likely to be highly non-linear, as shown in the a simplified version of the model in \ref{annex:sde}. For this reason it is hard to catch this trend with a linear regression.

The conclusion of these two regressions is that the time to convergence and the fluctuations around the equilibrium are influenced by the stochastic behaviour of the fitness function as well as its general shape. In physical terms, it means that the air traffic system as idealized by this model can be quite far from the equilibrium, due to 1) inadequate policies (the slope) 2) the general stochasticity of the system. Hence policy makers should carefully assess if any changes in policy is likely to have an impact due to the level of randomness of the system.

% ===================================================================== %
\section{Conclusions}
\label{section:conclusions}
% ===================================================================== %

In this paper, we have presented a stylized model of the allocation of flight plans. We used an agent-based model to simulate the behaviours of different air companies and the network manager. In the model, different types of air companies are competing for the best paths on the network of sectors and the best times of departure. Since the sectors are capacity-constrained, in some high traffic conditions the companies might be forced to choose suboptimal flight plans, according to their strategies or cost function. 

%In our model the infrastructure -- number of airports and network of sectors -- has a very predictable impact, opposed to the one of the increase of traffic. This effect is very well captured by a metric that we called ``overlap'', which depends heavily on the network structure. 

When different types of companies are competing on the same airspace, their relative satisfaction depends highly on the environment -- the airspace, the waves of departure -- but also the competition -- the fractions of different types of companies. In a nutshell, we find that the companies are performing better when they are competing against other types of companies, in a mechanism of ``niche'' leading to behaviours similar to those of the minority game \cite{minoritybook}.

As a consequence, it is possible to re-interpret the model as an evolutionary game, through the use of the difference of satisfactions as a fitness function which sets the capacity of a type of company to expand its business by having more flights in the future. In this framework the populations are the types of companies using the ``rerouting'' or ``shifting'' strategy. The study of the shape of the fitness function shows the existence of a stable equilibrium point for the mixing parameter for nearly every values of parameters. Interestingly, this equilibrium point is distinct from the point where a global satisfaction is optimal for the system as a whole. This indicates that the system left alone will not converge to the global optimum but to a different equilibrium point. 

In order to study more in details the real dynamics of the system around the point of equilibrium, we iterated the model with a reproduction rule mimicking the fact that higher satisfaction for an airline may be converted in better possibilities of expanding business. We found that the dynamical point of equilibrium is different from the one derived from the root of the fitness function (static equilibrium). This is a purely dynamical effect which is driving the point of equilibrium even further from the global optimum. Moreover, we found that both the convergence time to equilibrium and the fluctuations are highly dependent not only on the slope of the fitness function but also on the variance of the fitness function. 

These results have several policy implications. On one hand, the fact that the root of the fitness function is distinct from the global optimum means that the regulators should step in. Indeed, issuing well-designed regulations could help the system to have a point of equilibrium closer to the global optimum. On the other hand, the dynamical effects could blur the picture. Indeed, long time to convergence and high fluctuations combined with changing business conditions mean in practice that the system is always out of equilibrium. It is thus hard for the regulators to design incentives to drive the system to the optimum. A more precise setup and a more detailed calibration would be needed to definitely assert the potential consequences of regulations.

%Finally both are far from negligible, which might hinder the observation of the equilibrium in practice, due to changing environment.

The model presented here is an idealized version of the reality, a simple, yet phenomenologically rich, toy-model. It allows however to catch some high-level, emergent, phenomena that are inaccessible to more complicated ones due to the large number of parameters. The existence of a point of equilibrium, its behaviour in certain environments, and its dynamics  have certainly a scope broader than the present model. Moreover, the model is not really specific to the air traffic. In fact, it could be adapted to other situations, like packets propagation over the physical network of the internet, with minimal effort. As such, it can be viewed as a quite general model of transportation where entities need to send some material over a capacity-constrained network, thus competing for time and space.

Two possible directions for extension of the present work are the following. First, it is clear that a more detailed modelling could allow to draw some more precise conclusions about the present and future scenario in ATM. A first path has been made in this direction with another version of the model \cite{tactical}, based on navigation points instead of sectors. The model is also coupled to a tactical part, allowing to simulate the conflict resolution of traffic controllers. The code for this model is freely available \cite{code}. The second direction is toward model calibration. This is in general a challenging problem because data on strategic allocation are owned by companies and hardly available, especially when the details on many airlines are needed. A potential way of overcome this problem is through indirect calibration based on traffic data which contains the original flight plan and the last filled one. Data mining techniques could be useful to infer from this data unobservable parameters of our model and therefore to calibrate it.

\newpage

\appendix
%===================================================================== %
\section{Robustness of the model}
\label{annex:rob}
% ===================================================================== %

In this annex we test the robustness of the model to the simplifying  assumptions we made in the main text. Specifically, we consider two modifications of the baseline model. In the first one, instead of using a homogeneous network of airports, we use a more realistic one. In the second one, we use real data on airspace structure and airport network and choose the origin/destination pairs and the desired times accordingly.

\subsection{Scale free network for airports -- description}

It is well known that the distribution of degree in the airport network follows a power law \cite{guimera} and therefore is described by a scale free network. This means that few airports offer a large number of possible destinations, and many airports only a few possible ones. This is in contrast with the baseline model where we assumed that all airports are equivalent. In order to see if this feature changes our results, we decided to use a Barab\'{a}si-Albert type of network. This type of network generates a power law distribution for the degree (with exponent 3). In the simulations the air companies choose at random an origin/destination pair based on this airport network, i.e. they have a higher probability to be connected to high degree airports (hubs) rather than to low degree ones. The other properties of the model remain the same.

\subsection{Real network of sectors -- description}

For the second type of network, we use some traffic data (DDR) as well as some NEVAC files to have the definition of the sectors. All this data has been acquired during the course of SESAR funded WP-E project ELSA, ``EmpiricaLly grounded agent baSed models for the future ATM scenario''. More details about the data itself and the data acquirement process can be found in \cite{community}. 

\begin{figure}[htbp]
\begin{center}
\includegraphics[width=0.65\textwidth]{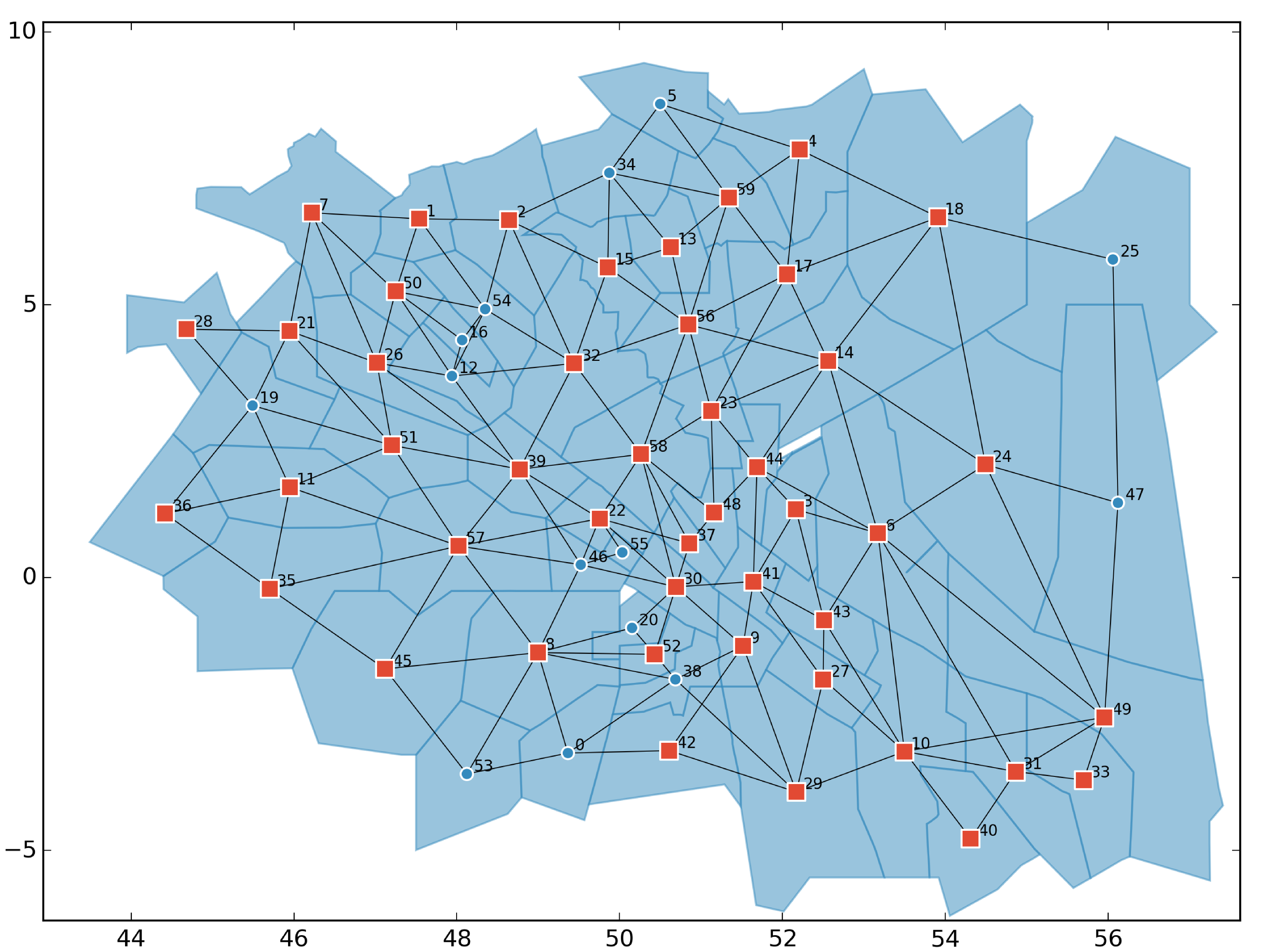}
\caption{Sector network used for the simulations based on traffic data. The sectors chosen are the ones which are the closest to the geographical center of the ECAC space (approximately around the Channel), at FL 350. The sectors are linked to each other if at least one flight goes from one to the other in the traffic data. The orange squares denote the sectors where there is at least one departure or one arrival. The numbers are arbitrary labels.}
\label{fig:map}
\end{center}
\end{figure}

For the purpose of testing the model, we used the data in the following way. First, we considered one day of traffic data, the 5th of June 2010. Since our model is in 2D of space, we needed to project the trajectories and have a unique tiling of the ECAC space. To this end, we selected FL 350 and considered only sectors at this altitude\footnote{In order to have a clean tiling, we needed to merge some overlapping sectors, probably active at different times during the day. Among 364 sectors, only 11 small sectors were slightly modified.}. From each flight trajectories we have extracted the path of sectors it actually followed. Then we selected only 60 sectors, in order to have results comparable with the previous ones (the generated networks have 60 sectors). We selected them by considering the most central sectors of the ECAC space (smallest distances between the center of the network and the center of the ECAC space). The resulting network of sectors is displayed in figure \ref{fig:map}.

The next step was to extract the times of crossing between sectors. For this, we computed from the traffic data the average time of flights going from one sector to another one. Because some flights were crossing only a very small portion of some sectors, we ended up having some strange crossing times (e.g. very small even for big sectors). So in order to keep the paths geographically sound, we computed the crossing times between sectors based only on their geographical distance, and tuned the average from all sectors to the data. 

The next step is to set the capacities for the sectors. Unfortunately, we do not have access to this kind of data. Instead, we decided to rely on the assumptions that the sectors were designed so that their capacity are only slightly bigger than the maximum traffic load. So we computed the maximum number of flights in each sector and fix it as the capacity. We are confident on the fact that the resulting capacities reflect at least the degree of heterogeneity of the capacities between sectors, even if not their absolute values. 

The final step is to extract from the data to possible origin/destination pairs and the desired times of departure. For the pairs, we simply recorded the first and last sectors crossed by the flights in the area. For the departure times, we assumed that the last filled flight plan available in the data constituted a good approximation of the desired departure times. 

We then ran some simulations, changing the number of flights $N_f$ and the mixing parameter $f_S$. For each set of parameters, we produced 100 simulations. In each of them, each air company first picks at random an origin/destination pair from the available ones. Then it picks at random a desired departure time among the available ones. Finally, it picks a strategy (type S or R) with a probability $f_S$.

\subsection{Results}

\begin{figure}[htbp]
\begin{center}
\includegraphics[width=0.5\textwidth]{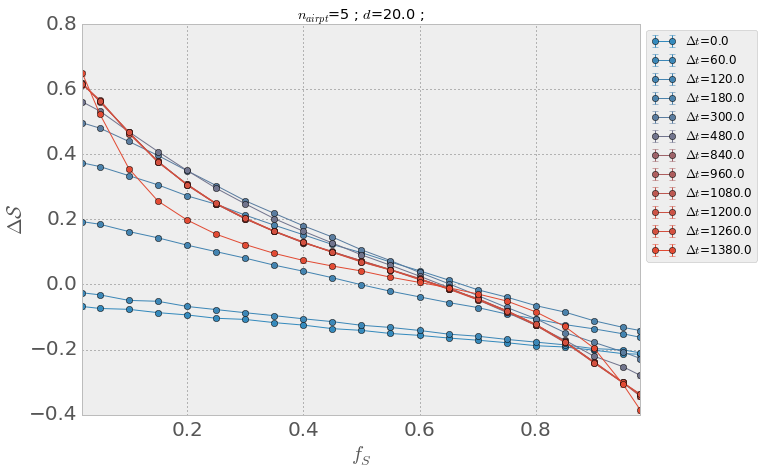}
\includegraphics[width=0.46\textwidth]{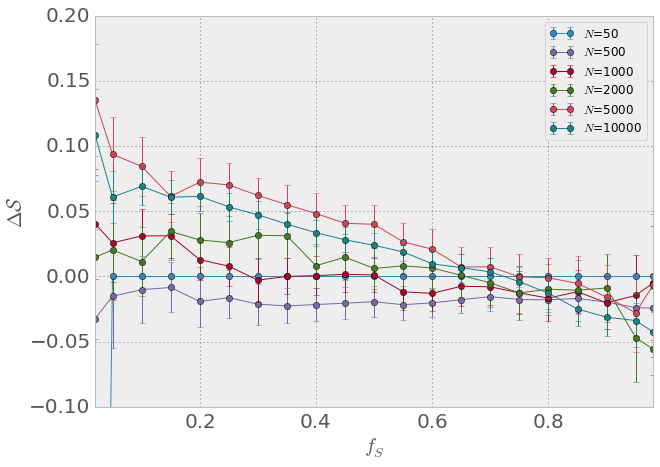}
\caption{Simulation results of the two robustness checks of the model. Left: difference of satisfaction $\Delta S$ between the two types of companies as a function of the mixing parameters $f_S$ for different values of $\Delta t$ for an artificially generated  Barab\'{a}si-Albert airport network. Right: difference of satisfaction as a function of the mixing parameter for different numbers of flights $N$, in the case where the network is constructed from real traffic data.}
\label{fig:rob}
\end{center}
\end{figure}

Figure \ref{fig:rob} shows the most important result from the two previous procedures, namely the evolution of the difference of satisfaction as a function of the mixing parameter $f_S$. For the artificial scale free network of sectors we varied the time $\Delta t$ between waves, while in the simulations on the real network the wave structure is fixed by the real data and therefore we considered different number of flights $N$. 

Figure \ref{fig:rob} should be compared with figure \ref{fig:diff_S}. We observe that the general inverse relation between $\Delta S$ and $f_S$ is the same as in the baseline model, i.e. companies are usually performing better when they are competing against large populations of the other type of company. The real case on the right is just a bit different in the magnitude of the change of satisfaction. It seems that the effect of the mixing parameter is weaker in this case. However for large traffic (high $N$) the inverse relation is very clear.

% ===================================================================== %
\section{Analytical derivation of equilibrium for infinite and finite populations}
\label{annex:sde}
% ===================================================================== %

The standard replicator model with two populations \cite{nowak} fixes the total population size and describes the dynamics of the fraction of population of a given type (for example companies S). If $x(t)\in [0,1]$ is the fraction of population of a given type at time $t$ (the other being $1-x(t)$), the dynamics in the infinite population case is 
$$
\dot{x} = f(x) x(1-x),
$$
where $f(x)$ is the difference of the fitness between the two populations and $\dot{x}$ is the derivative of $x$ with respect to time. In this expression, the fitness function is a deterministic function of one variable, $x$. Clearly the equilibrium points of this dynamical system are $x=0$, $x=1$, and the roots (if any) of $f$, i.e. the points where $f(x_{eq})=0$.

However, for finite populations, the fitness can depend on the exact realization of the model, because of different origin/destination distributions for instance. In order to take this into account, we substitute $f(x)$ by $f(x) + \sigma \eta$, where $\eta$ is a Gaussian white noise with mean zero and variance 1. The noise term $\eta$ has a variance $\sigma^2$ which goes to zero when the population size goes to infinity.
%In principle $\sigma$ could depend also on $x$, but we take the simplified case where it is a constant. 
The equation is now a nonlinear Langevin equation \cite{gardiner}. In order to solve the equation, we linearize it around the point of equilibrium. With $\tilde{x}_t = x_t - x_{eq}$ and $f(x) = \gamma \xx_t$, where $\gamma$ ($<0$) is the derivative of the fitness at $x_{eq}$, the equation becomes:
$$
\dot{\xx}_t =  \gamma x_{eq} (1-x_{eq}) \xx_t + (x_{eq}(1-x_{eq}) + (1-2x_{eq})\xx_t)\sigma \eta_t.
$$
Since the noise term is multiplicative, in order to trasform it in a Stochastic Differential Equation (SDE), we assume Stratonovich integration \cite{gardiner} and we obtain
$$
d\xx_t = a \xx_t dt +  (d + b\xx_t)\sigma dW_t, ~~~~(Stratonovich)
$$
where $W_t$ is a Wiener process and where $a =  \gamma x_{eq} (1-x_{eq})$,  $d = x_{eq}(1-x_{eq})$, and  $b =  (1-2x_{eq})$.

We now pass to the Ito formalism\footnote{Given $dX_t=\alpha_t dt+\beta_t dW_t$ in Stratonovich sense, the corresponding Ito equation is $dX_t=(\alpha_t+\frac{1}{2}\beta_t \partial_x \beta_t)dt+\beta_t dW_t)$ \cite{gardiner}. } obtaining
\begin{equation}\label{eq_ito}
d\xx_t =A(B-\xx_t)dt+ C\sigma dW_t~~~~~~~(Ito)
\end{equation}
where 
\begin{eqnarray*}
&A&=-\gamma x_{eq} (1-x_{eq})-\frac{(1-2x_{eq})^2}{2}\sigma^2 \\
&B&=\frac{x_{eq}(1-x_{eq})(1-2x_{eq})\sigma^2}{-2\gamma x_{eq}(1-x_{eq})-(1-2x_{eq})^2\sigma^2}\\
&C&= x_{eq}(1-x_{eq})+(1-2x_{eq})\xx_t
\end{eqnarray*}

Equation \ref{eq_ito} is a linear Ito SDE which can be easily solved. The linear drift term $A(B-\xx_t)$ tells us that the dynamics of $\xx_t$ is mean reverting around the position $B$ at an exponential rate $(-A)^{-1}$, when $A<0$. Framed in our problem this fact has two implications:
\begin{itemize}
\item The dynamical equilibrium point is $x'_{eq}=x_{eq}+B$. Since in our model $x_{eq}>1/2$, it is  $x'_{eq}>x_{eq}$ i.e. the new equilibrium has a larger fraction of S companies with respect to the value obtained from the infinite population case.
\item The speed of convergence to the equilibrium is $(-A^{-1})$ which is larger than the zero noise case $(-\gamma x_{eq} (1-x_{eq}))^{-1}$. This implies that convergence is reached more slowly than in the infinite population case. 
\end{itemize}

Therefore, as observed in simulations, the equilibrium in finite populations favors even more S companies and it is reached at a slower rate that the one predicted by the slope of the fitness at the equilibrium point. Finally, as expected, when the population size increases, $\sigma \to 0$ and the infinite dimensional solutions are recovered.

\end{document}